\documentclass[amssymb,useAMS,prd,aps,amsmath,nofootinbib,superscriptaddress,twocolumn]{revtex4}

\usepackage{graphicx}
\usepackage{amssymb}
\usepackage{wasysym}
\usepackage[caption=false]{subfig}
\usepackage{array, xcolor}
{

\usepackage[colorlinks=true,linkcolor=blue,citecolor=blue,urlcolor=black]{hyperref}

\begin{document}
\title{Fitting the annual modulation in DAMA with neutrons from muons and neutrinos }
\author{Jonathan H. Davis}
\affiliation{Institute for Particle Physics Phenomenology, Durham University, Durham, DH1 3LE, United Kingdom \\ {\smallskip \tt  \href{mailto:j.h.davis@durham.ac.uk}{j.h.davis@durham.ac.uk}\smallskip}}

\begin{abstract}
The DAMA/LIBRA experiment searches for evidence of Dark Matter scattering off nuclei. Data from DAMA show $9.2\sigma$ evidence for an annual modulation, consistent with Dark Matter having a cross section around $2 \cdot 10^{-40}$ cm$^2$. However this is excluded by other Direct Detection experiments. We propose an alternative source of annual modulation in the form of neutrons, which have been liberated from material surrounding the detector by a combination of $^8$B solar neutrinos and atmospheric muons. The phase of the muon modulation lags $30$~days behind the data, however we show that adding the modulated neutrino component shifts the phase of the combined signal forward. 
In addition we estimate that neutrinos and muons need $\sim 1000$~m$^3$ of scattering material in order to generate enough neutrons to constitute the signal. With current data our model gives as good a fit as Dark Matter and we discuss prospects for future experiments to discriminate between the two.
\end{abstract}

\maketitle
\section{Introduction}
The DAMA/LIBRA (and formerly DAMA/NaI) experiment searches for keV-energy nuclear recoils, potentially arising from Dark Matter (DM) originating in the galactic halo~\cite{Bernabei:2013xsa,Bernabei:2013cfa,Bernabei:2008yi}. It operates with approximately $250$~kg of NaI, located deep underground at Gran Sasso. The DAMA/LIBRA collaboration claim to have observed a temporal variation in the rate of observed events with a period of roughly one year and at a significance of $9.2\sigma$. This annual modulation is of the order of $2\%$ and is approximately sinusoidal, with a maximum in late May. 

Such an annual modulation is consistent with Dark Matter scattering off nuclei inside the detector, since the relative direction of the incoming DM varies over the course of the year and peaks around June 2nd, similarly to the DAMA data.

However the annual modulation observed by DAMA requires a cross section of interaction between DM and nucleons of $\sigma \approx 2 \cdot 10^{-40}$ cm$^2$~\cite{Schwetz:2011xm} for a mass $m \approx 10$~GeV (and elastic scattering). Unfortunately the values of $m$ and $\sigma$ favoured by DAMA are excluded by other Direct Detection experiments such as CDEX~\cite{Yue:2014qdu}, CDMS-II~\cite{Ahmed:2010wy,2012arXiv1203.1309C}, EDELWEISS-II~\cite{Armengaud:2012pfa}, LUX~\cite{Akerib:2013tjd}, SuperCDMS~\cite{Agnese:2014aze}, XENON10~\cite{Angle:2011th} and XENON100~\cite{Aprile:2012_225}. This motivates alternative explanations for the DAMA signal.

One alternative source of an annual modulation is cosmic ray muons \cite{Blum:2011jf,Ralston:2010bd}, whose flux is correlated with the temperature of the atmosphere~\cite{D'Angelo:2011fs,FernandezMartinez:2012wd}. The DAMA signal is then explained as being made up of neutrons which have been liberated by muons interacting in the rock surrounding the detector~\cite{Blum:2011jf,Ralston:2010bd}. However, although the period is consistent with the DAMA data, the phase of the muon-induced neutron signal is not i.e. the muon flux peaks roughly 30 days too late~\cite{Blum:2011jf,FernandezMartinez:2012wd,Chang:2011eb,Bernabei:2012wp}. Hence the muon signal is incompatible with DAMA at $5.2\sigma$~\cite{FernandezMartinez:2012wd}. 

In this letter we propose a solution in the form of an additional source of neutrons, generated by $^8$B solar neutrinos interacting in the rock or shielding surrounding the DAMA detector. Crucially the solar neutrino flux varies annually and peaks around January 4th, due the the changing distance between the Earth and Sun. We show in section~\ref{sec:time} that when combined with the neutrons from cosmic muons the phase of the signal can be shifted forward by $\sim 30$~days relative to muons-alone, resulting in a fit to the data as good as that from Dark Matter. 

This shift relies upon a degree of cancellation between the two modulated rates and so requires the neutron flux from muons $R_{\mu}$ and $^8$B neutrinos $R_{\nu}$ to be of a similar size. In section~\ref{sec:rates} we demonstrate that this is in fact the case for the Gran Sasso lab where $R_{\nu} / R_{\mu} \sim 0.1$, as the large neutrino flux compensates for its small cross section relative to muons. In section~\ref{sec:higher_modes} we discuss methods of discriminating our model from Dark Matter using for example higher-order modes and we conclude in section~\ref{sec:conc}.

\section{Annual modulation of neutrinos, muons and dark matter \label{sec:time}}
\begin{figure*}[t]
\centering
\subfloat{ \includegraphics[width=0.99\textwidth]{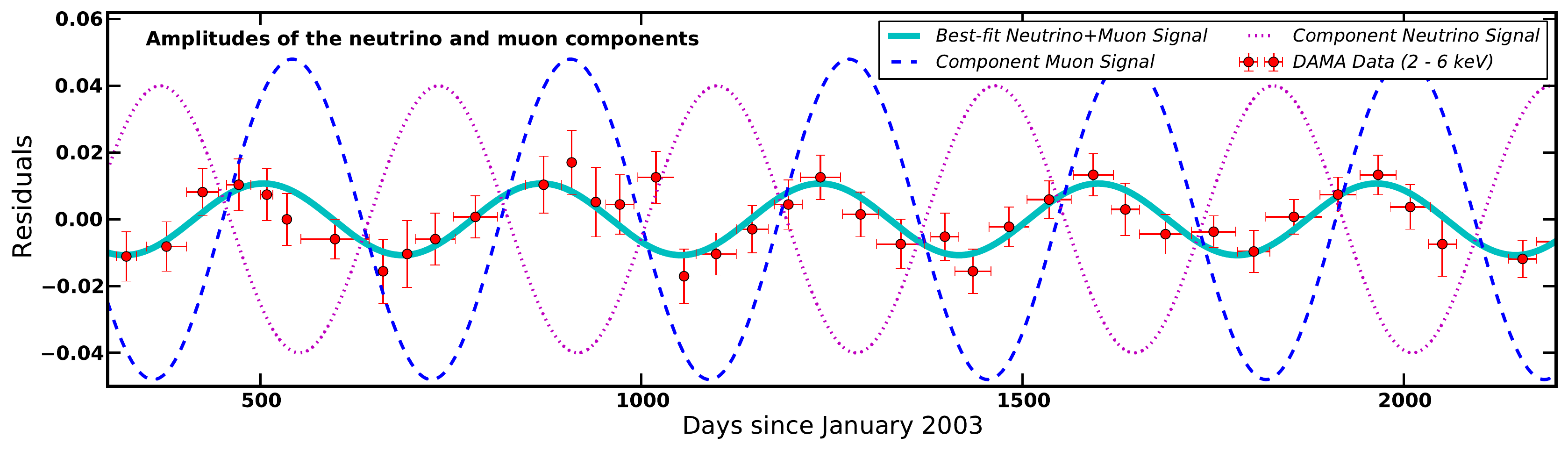}} \\
\caption{The DAMA signal is composed of neutrons liberated in the material surrounding the detector by both solar neutrinos (dotted) and atmospheric muons (dashed). Both components have fixed phases, with only their amplitudes as free parameters. Individually neither of these has the correct phase to fit the data, however in combination the fit quality is excellent.}
\label{fig:all_neutrons}
\end{figure*}

\begin{figure}[t]
\centering
\subfloat{\includegraphics[width=0.49\textwidth]{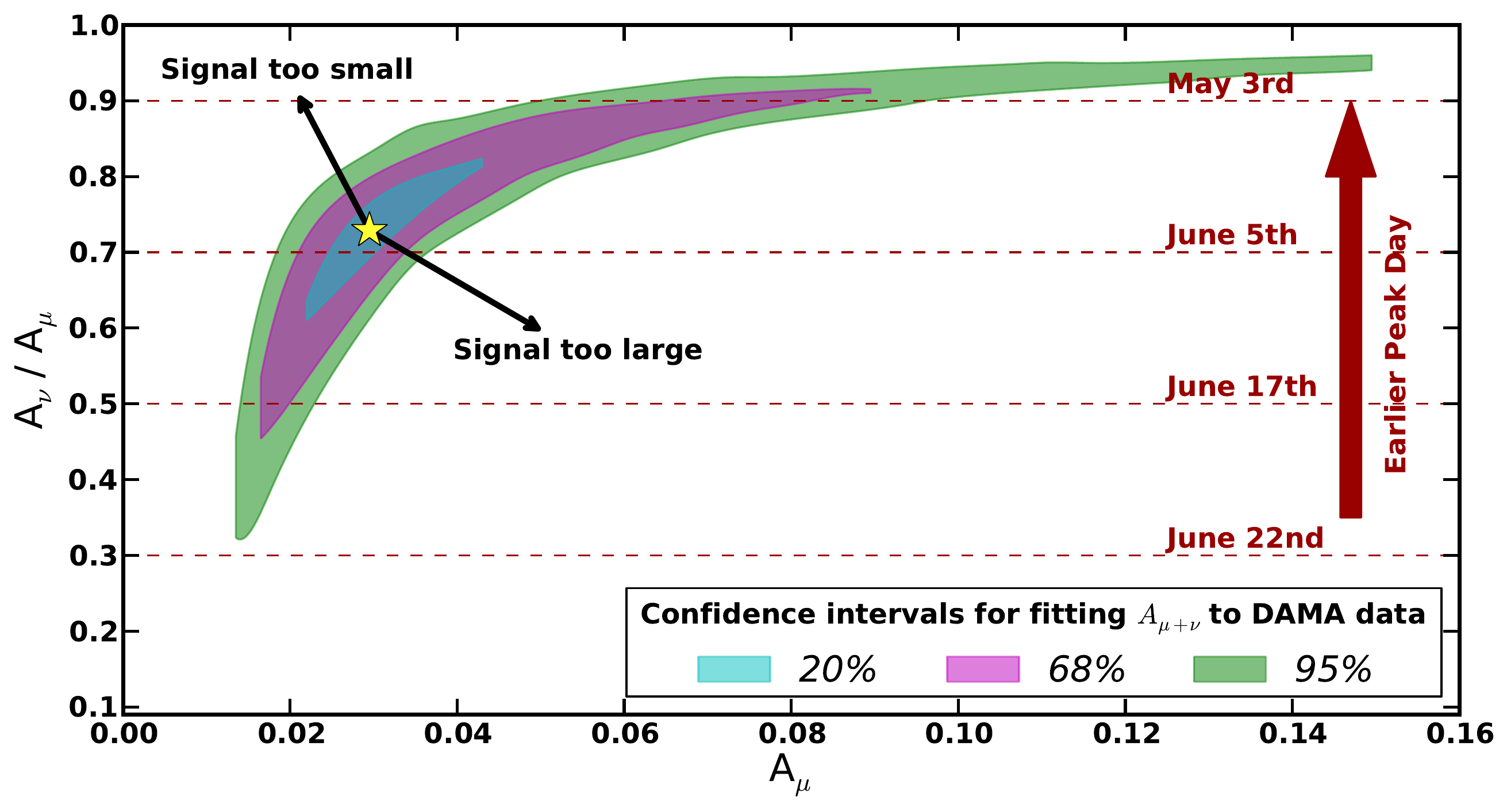}}
\caption{Contours (and best-fit point $\star$) of the modulation residuals for the muon $A_{\mu}$ and neutrino $A_{\nu}$ induced neutron signal in equation \eqref{eqn:mu_nu}, for the case where the phases are marginalised over. Shown also are approximate values for the day where the signal peaks for selected values of $A_{\nu}/A_{\mu}$.}
\label{fig:contours}
\end{figure}

\begin{figure*}[t]
\centering
\subfloat{ \includegraphics[width=0.99\textwidth]{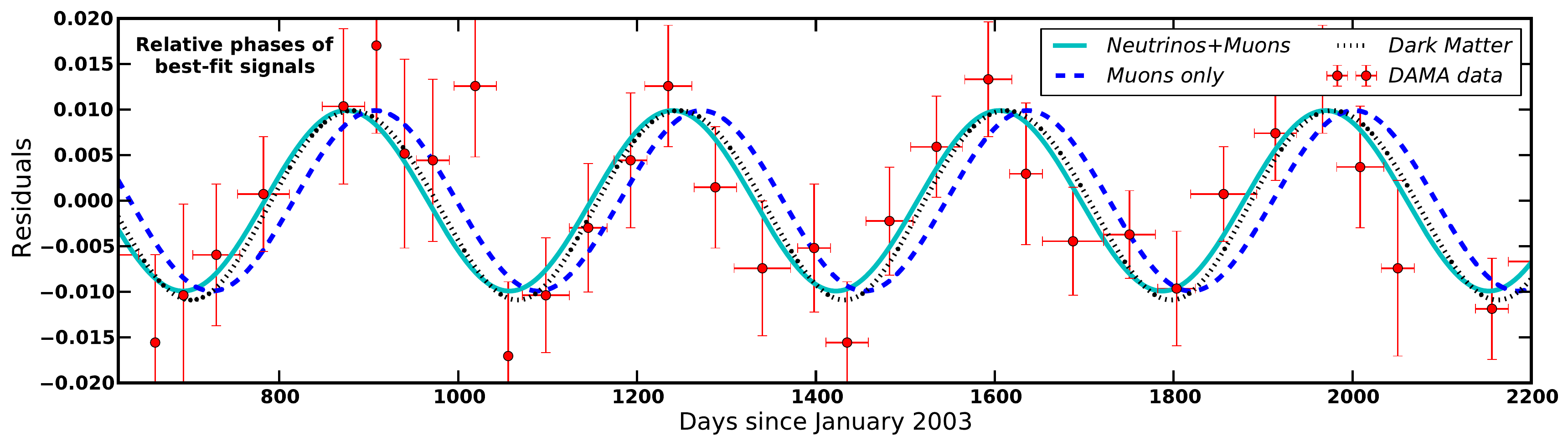}} 
\caption{Comparison of  models for the DAMA data. The model proposed in this letter is shown as the solid cyan line, composed of neutrons produced by solar neutrinos and atmospheric muons~(with fixed phases $(\phi_{\nu},\phi_{\mu}) = (3,179)$~days).  Adding the solar neutrino contribution to that from muons shifts the phase forward by $\sim 30$~days, markedly improving the fit to the data.}
\label{fig:all_neutrons_2}
\end{figure*}

In this section we introduce the cosmic muon and neutrino signals, with the aim of fitting their time-variation to the combined data from DAMA/NaI and DAMA/LIBRA, and comparing the fit to that from Dark Matter. We use the full data-set over 13 years with a 1.17 ton year exposure in the $2$~keV to $6$~keV bin~\cite{Bernabei:2013xsa,Bernabei:2013cfa}.

The solar neutrino flux at Earth depends on the distance between the Earth and Sun according to an inverse-square relation. Since the Earth's orbit is slightly eccentric, the distance between the Earth and Sun varies with a period of a year, and therefore so does the neutrino flux. Hence the flux is given by the expression~\cite{Bellini:2013lnn},
\begin{equation}
\Phi_{\nu} = \frac{\mathcal{R}}{4 \pi r^2(t)} \approx \frac{\mathcal{R}}{4 \pi r_0^2} \left[ 1 + 2 \epsilon \mathrm{cos} \left(\frac{2 \pi (t - \phi_{\nu})}{T_{\nu}}\right)  \right] ,
\label{eqn:nu}
\end{equation}
where $\mathcal{R}$ is the neutrino production rate in the Sun, $t$ is the time from January 1st, $r(t)$ is the time-dependent distance between the Earth and Sun, $r_0$ is the average distance, $\epsilon = 0.01671$ is the orbital eccentricity, $T_{\nu}$ is the period and $\phi_{\nu}$ is the phase. The Earth is closest to the Sun around January 4th (implying $\phi_{\nu} = 3$~days).

This has been confirmed experimentally.
Measurements from Borexino~\cite{Bellini:2013lnn} for $^7$Be neutrinos imply a period of $T_{\nu} = 1.01 \pm 0.07$~years and a phase of $\phi_{\nu} = 11.0 \pm 4.0$~days. Additionally the flux of $^8$B solar neutrinos has been observed by Super-Kamiokande~\cite{PhysRevD.78.032002} to be consistent with variation in the Earth-Sun distance.

The muons  originate from the decay of cosmic ray particles in the stratosphere. These parent particles can also collide with the air, with more collisions leading to fewer muons being produced. 
In the winter the rate of collisions is largest and so the muon flux is lowest. Hence the muon flux is correlated with the temperature of the atmosphere~\cite{Ambrosio:1997tc}, giving the expression
\begin{equation}
\Phi_{\mu} \approx \Phi_{\mu}^0 + \Delta \Phi_{\mu} \mathrm{cos} \left({2 \pi (t - \phi_{\mu})}/{T_{\mu}}\right),
\label{eqn:mu}
\end{equation}
where $\Phi_{\mu}^0$ is the average cosmic muon flux, $T_{\mu}$ is the period and $\phi_{\mu}$ the phase.

Measurements of muons by Borexino~\cite{D'Angelo:2011fs} imply that $T_{\mu} = 366 \pm 3$~days, $\Delta \Phi_{\mu} / \Phi_{\mu}^0  = 0.0129 \pm 0.0007$ and $\phi_{\mu} = 179 \pm 6$~days. Hence the phase and period are consistent with an annual modulation of muons peaked on approximately June 21st.

We seek to explain the DAMA annual modulation using a combination of the neutrino and muon signals. The signal itself is due to neutrons, which are liberated in the rock or shielding by the neutrinos and muons. 
Our signal therefore takes the form~of,
\begin{equation}
A_{\mu+\nu} = A_{\nu}  \mathrm{cos} \left(\omega (t - \phi_{\nu})\right) +  A_{\mu}  \mathrm{cos} \left(\omega (t - \phi_{\mu})\right),
\label{eqn:mu_nu} 
\end{equation}
where $\omega = 2\pi / T$. The amplitudes $A_{\mu}$ and $A_{\nu}$ correspond to the modulation `residual' which is the relative deviation of the event rate from the time-average.

Since there exist no direct measurements of the modulation residuals for low-energy muon and neutrino induced neutrons we leave their amplitudes as free parameters.
Indeed we perform two separate fits of $A_{\mu+\nu}$ to DAMA data: for the first we leave the amplitudes $A_{\nu}$ and $A_{\mu}$ as free variables and fix the other parameters to be $(T,\phi_{\nu},\phi_{\mu}) = (365,3,179)$~days. For the second we marginalise over $\phi_{\nu}$ and $\phi_{\mu}$ with Gaussian priors centred on the best-fit values from Borexino. As such the phases are not free parameters, but are fixed \emph{a priori}.

Shown in figure~\ref{fig:all_neutrons} is the result of our first fit to DAMA data, with the phases held fixed. We obtain best-fit amplitudes of $A_{\nu} = 0.039$ and $A_{\mu} = 0.047$ with $\chi^2 = 66.74$. 
For the second fit we obtain best-fit amplitudes of $A_{\nu} = 0.022$ and $A_{\mu} = 0.030$.

We show best-fit contours for the marginalised fit in figure~\ref{fig:contours}. 
The signal fits well for a wide-range of amplitudes and the best-fit is given when the peak day matches that of the DAMA data. This depends on the phases of the two signals (which are known \emph{a priori}) and their relative amplitudes, with larger values of $A_{\nu}/A_{\mu}$ shifting the peak day to earlier times. 
Additionally the fit gets worse whenever the amplitude of the combined signal is either too small or too large, as indicated by the arrows.

We now compare our fit to that from a Dark Matter signal. 
The differential rate of Dark Matter interactions with nuclei takes the form of
\begin{equation}
\frac{\mathrm{d}R}{\mathrm{d}E} = \frac{\rho_{\mathrm{DM}}}{m_{\mathrm{N}} m} \int \mathrm{d}^3 v \frac{\mathrm{d}\sigma}{\mathrm{d}E} v f(\mathbf{v} + \mathbf{v}_E(t)),
\end{equation}
where $\rho_{\mathrm{DM}}$ is the DM density, $m_{\mathrm{N}}$ is the mass of the target nucleus and $\mathrm{d}\sigma / \mathrm{d}E$ is the differential cross section. 

The integral is over the galactic DM velocity distribution $f(\mathbf{v})$ boosted into the Earth's rest-frame by $\mathbf{v}_E(t)$. The time-dependence enters via this term, expressed as  $\mathbf{v}_E(t) = \mathbf{v}_0 + \mathbf{v}_{\mathrm{pec}} + \mathbf{u}_E(t)$, where $\mathbf{v}_0 = (0,220,0)$~kms$^{-1}$ and the peculiar velocity $\mathbf{v}_{\mathrm{pec}} = (11.1 \pm 1.2,12.2 \pm 2.0,7.3 \pm 0.6)$~kms$^{-1}$~\cite{2010MNRAS.403.1829S}. For the relative velocity between the Earth and the Sun $\mathbf{u}_E(t)$ we use the expression from~\cite{McCabe:2013kea}. We assume a Maxwell-Boltzmann distribution for $f(\mathbf{v})$. Allowing the amplitude to vary freely we obtain a best-fit chi-square of~$\chi^2 = 69.76$.

We show in figure~\ref{fig:all_neutrons_2} the neutrino+muon signal from our first fit (with $A_{\nu} = 0.039$ and $A_{\mu} = 0.047$) compared with a Dark Matter signal and the best-fit signal from muons-alone.
The neutrino+muon and Dark Matter signals are very close together in phase and both fit well to the DAMA data. As expected the muon-only model provides the worst fit as it has a phase which lags $\sim 30$~days behind the data.  This is confirmed by the $\chi^2$ values, which we show in table~\ref{table_chi2}.

We present two additional metrics in table~\ref{table_chi2}, which account for the different numbers of free parameters. For the Akaike Information Criterion~\cite{1100705} the neutrino+muon model gives the best fit and for the  Bayesian Information Criterion Dark Matter has the lowest value, but only by a difference of $\Delta$BIC$ = 1.36$ which is not significant.
We conclude that our neutrino+muon model fits as well to the DAMA modulation as a Dark Matter signal.

\begin{table}[t]
\begin{center}
\begin{tabular}{ c || c | c | c }
 & \normalsize{$\chi^2$} & \normalsize{AIC} & \normalsize{BIC} \\
\hline
\normalsize{Muons and Neutrinos} & \normalsize{66.74} & \normalsize{70.74} & \normalsize{75.50} \\
 \normalsize{Dark Matter} & \normalsize{69.76} & \normalsize{71.76} & \normalsize{74.14} \\
  \normalsize{Muons-only} & \normalsize{90.39} & \normalsize{92.39} & \normalsize{94.77} \\
\end{tabular}
\end{center}
\caption{Compatibility of three models with DAMA data, where $\mathrm{AIC} = \chi^2 + 2k$ and $\mathrm{BIC} = \chi^2 + k \mathrm{ln} \, n$, with $k$ as the number of parameters and $n = 80$ the number of data-points.}
\label{table_chi2}
\end{table}

\section{Rates of cosmogenic neutrons \label{sec:rates}}

We have modelled the DAMA annual modulation using neutrons produced by solar neutrinos and atmospheric muons. Indeed the DAMA events can not be due \emph{directly} to muon or neutrino scattering, due to statistical arguments for the former~\cite{Blum:2011jf} and too small a rate for the latter~\cite{Bernabei199545}.
In this section we discuss whether these muons and neutrinos can produce enough neutrons to constitute the DAMA signal.

Muons produce neutrons via scattering in either the rock or potentially the lead shielding around the detector~\cite{Blum:2011jf,Ralston:2010bd,Araujo:2004rv}. 
Likewise neutrons from neutrino neutral-current scattering have been proposed as a detection method for supernovae neutrinos using $^9$Be, $^{23}$Na, $^{35}$Cl, $^{56}$Fe and $^{208}$Pb targets~\cite{Fuller:1998kb,PhysRevD.50.720,2011JCAP...10..019V,Kolbe:2003ys}. For $^{208}$Pb the neutron emission threshold for the neutrino is $E_{\nu} > 7.37$~MeV~\cite{Fuller:1998kb} and so $^8$B solar neutrinos could stimulate neutron spallation, since these have energies up to $14$~MeV~\cite{PhysRevD.78.032002,Bellini:2013lnn}.

We now calculate the amount of target needed for cosmogenic neutrons to explain the DAMA signal. We estimate the rate of neutrons using $R \sim \Phi \sigma n V$, where $\Phi$ is the flux, $\sigma$ is the interaction cross section, $n$ is the number density of the target and $V$ is its volume.

For $^8$B solar neutrinos the flux is of the order $\Phi_{\nu} \sim 10^6$~cm$^{-2}$s$^{-1}$~\cite{PhysRevD.78.032002}. Assuming a $^{208}$Pb target the cross section for neutrino-induced neutron spallation is $\sigma \sim 10^{-41}$~cm$^2$~\cite{Fuller:1998kb}. Hence the rate of neutrino-induced neutron emission is of the order $R_{\nu} \sim 10^{-35} n V$~neutrons/sec.
For muons we assume a flux at the Gran Sasso lab of $\Phi_{\mu} \sim 10^{-8}$~cm$^{-2}$s$^{-1}$ and a cross section for neutron production $\sigma \sim 10^{-26}$~cm$^2$~\cite{Araujo:2004rv}. This gives a muon-induced neutron rate of $R_{\mu} \sim 10^{-34} n V$~neutrons/sec. Hence our estimates imply $R_{\nu} / R_{\mu} \sim 0.1$ which is encouraging given that we required for the modulation residuals $A_{\nu} / A_{\mu} \approx 0.5$ to provide a good fit to DAMA data.

Taking the number density to be $n = 10^{29}$~m$^{-3}$ a volume of $V \sim 1000$~m$^3$ is enough to generate $\sim 100$~neutrons per day, which is similar to the rate observed in DAMA. 
For the muon-induced neutrons the mean free path (MFP) is $\lambda \approx 2.6$~m~\cite{Barker:2012nb}. Hence  we estimate the effective volume over which these neutrons are produced and still reach the detector to be $V_{\mathrm{eff}} = 4 \pi \int \mathrm{d} r \, r^2 \exp[-r / \lambda] \approx 450$~m$^3$, which is close to volume $V$ needed to explain the DAMA signal. The neutrino-induced neutrons will be of lower energies resulting in a shorter MFP and a smaller $V_{\mathrm{eff}}$. However we have neglected the production of multiple neutrons per neutrino and resonances in the cross section for neutron production, which may compensate for the smaller volume.

%This is of a similar size to the volume $V$ which we need to explain DAMA and so our estimate for the neutron rate seems reasonable.

We have yet to comment on why only DAMA  (and perhaps CoGeNT~\cite{Aalseth:2014eft,Davis:2014bla}) sees a modulation signal~\cite{2012arXiv1203.1309C}. This could be due to a combination of several factors, most notably shielding and thresholds. For the former other experiments employ different, and possibly stronger, neutron shields~\cite{Akerib:2014rda,2014arXiv1406.2374A}. The geometry of the shielding may also be important: for example at KIMS the polyethylene shield is between the lead shield and the detector~\cite{kims_pres}.

 For the latter it is known that muons produce neutrons with a spectrum which rises at low energies, and so the majority of the muon-induced neutrons have kinetic energy of the order $ 10 - 100$~keV~\cite{Araujo:2004rv}.
The neutrino-induced spectrum will be similar, but also includes a population of low-energy neutrons from neutrino scatters near the detector.
Hence if the neutrons are scattering off Na in DAMA then the recoil energies fall into the $2 - 6$~keV bin. However for heavier targets such as xenon or germanium the recoil energies would likely be below threshold.

\section{Higher-order modes and future tests \label{sec:higher_modes}}
\begin{figure}[t]
\centering
\subfloat{\includegraphics[width=0.49\textwidth,trim=0 0 0 0,clip=True]{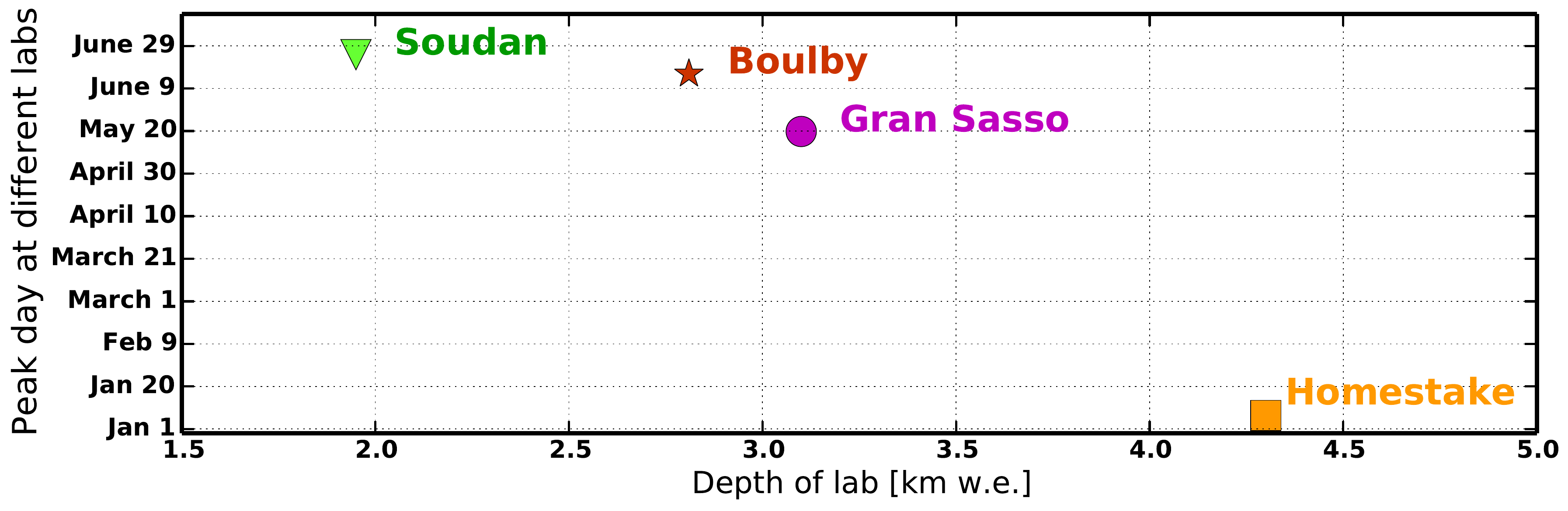}}
\caption{Approximate peak day of the neutrino+muon signal at four different labs. Deeper labs have a lower muon flux~\cite{Mei:2005gm} and so a phase closer to that of the solar neutrinos. 
}
\label{fig:labs}
\end{figure}

Based on annual modulation alone the muon+neutrino and Dark Matter models provide equally valid fits to the DAMA data. In this section we discuss methods for discriminating between these two models.

The Sun goes through cycles of activity with a period of approximately 11 years.  Indeed atmospheric muons possess a signifiant additional modulated component with a period of $10.7 \pm 0.3$~years~\cite{FernandezMartinez:2012wd,Chang:2011eb}. 
However no correlation between solar activity and $^8$B solar neutrinos has been observed by Super-Kamiokande~\cite{PhysRevD.78.032002} or for $^7$Be neutrinos in Borexino~\cite{Bellini:2013lnn} (though there may be evidence of a quasi-biennial mode~\cite{Sakurai:1981}). Hence it is not possible for the two signals to interfere at the 11 year mode as they do for the annual mode (i.e. $A_{\nu}/A_{\mu}|_{T = 11~\mathrm{year}} < A_{\nu}/A_{\mu}|_{T = 1~\mathrm{year}}$) and so a lack of power for this harmonic in the data would exclude our model.

%Unfortunately it is difficult to look for such an 11 year mode in the DAMA data-set as the collaboration subtract the average residual on a yearly basis, suppressing power at long periods~\cite{Chang:2011eb}. 
We do not have the full DAMA data-set, however it is possible to partially reconstruct this information using the annual averages~\cite{Bernabei:2013cfa,Chang:2011eb}. 
Indeed the annually-averaged rates appear consistent with the $\sim 1\%$ modulation expected from the 11 year mode.
However a statistical analysis of this data is difficult, since the earliest data comes from DAMA/NaI (with larger error bars) while the later part comes from DAMA/LIBRA~\cite{Bernabei:2013cfa}.

There are further tests which may be able to distinguish our model from Dark Matter.
For example, in section~\ref{sec:time} we used only the 2~keV to 6~keV energy-bin. However the DAMA collaboration have shown that the best-fit phase shifts forwards by $\sim 10$~days when fitting to events with energies between  2~keV and 4~keV~\cite{Bernabei:2013xsa}.   
Since $A_{\nu}/A_{\mu}$ could increase with lower energy, such a shift may be expected from our combined model.

Additionally, our model predicts a modulation in double-scatter neutrons. However this will likely be washed out by other more numerous unmodulated double-scatter events, such as gammas.

%, since DAMA can not effectively distinguish between sources (e.g. nuclear vs. electronic recoils).

Finally future experiments such as DM-Ice~\cite{Cherwinka:2014xta}, KIMS~\cite{2012PhRvL.108r1301K}, SABRE~\cite{sabre_pres} or ANAIS~\cite{Amare:2013lca} will be able to exploit a location-dependent phase change e.g. due to the depreciation of muon flux with depth. We show in figure~\ref{fig:labs} the expected peak day of the neutrino+muon model at four labs. For example if $A_{\nu} / A_{\mu} = 0.7$ at Gran Sasso it will be $0.44$ at Boulby since the muon flux is $1.6$ times larger~\cite{Mei:2005gm}, leading to a peak day $\sim 20$~days later. 

%Hence the phase (and amplitude) of a second modulation result at e.g. DM-Ice~\cite{Cherwinka:2014xta}, KIMS~\cite{2012PhRvL.108r1301K}, SABRE~\cite{sabre_pres} or ANAIS~\cite{Amare:2013lca} would strongly constrain our model.

% It would also be interesting if these experiments could vary their neutron shielding, in order to test a neutron origin for the DAMA signal.

\section{Conclusion \label{sec:conc}}
In this letter we have proposed a new model for the DAMA annual modulation, which is a sum of two annually modulating components with different phases. More specifically the events are composed of neutrons, which are liberated in the material surrounding the detector by a combination of $^8$B solar neutrinos and atmospheric muons. The model is shown in figure~\ref{fig:all_neutrons}. 

The muons alone can not explain the DAMA annual modulation, as has been remarked upon before~\cite{Blum:2011jf,FernandezMartinez:2012wd,Chang:2011eb}, since they peak approximately $\sim 30$~days too late. Inclusion of the solar neutrinos solves this issue as they also modulate and peak around January 4th, effectively shifting the phase of the combined model forward. This is shown in figure~\ref{fig:all_neutrons_2}.
Due to this phase shift we found that our model fits as well to the DAMA annual modulation as Dark Matter.

We have shown that both the muon and neutrino signals can produce enough neutrons provided they scatter in a volume approximately $\sim 1000$~m$^3$ in size around the DAMA detector. 
A detailed Monte Carlo simulation of the detector is required in order to verify if our determination of the neutron rate is realistic.

However this degeneracy between the muon+neutrino model and Dark Matter extends only to modulation with a period of one-year.  A search for an 11 year mode or an energy-dependent phase may break the degeneracy and future experiments will additionally be able to study if the modulation phase varies with location.

Hence it is premature to disregard cosmogenic neutrons as an explanation for the DAMA modulation based on the phase, and our model presents a testable alternative for future experiments aiming to look for an annual modulation due to Dark Matter.

\section*{Acknowledgements}
The author would like to thank Henrique Araujo, Celine Boehm, Brian Feldstein, Chris McCabe and Ryan Wilkinson for helpful comments and discussions and the STFC for their generous financial support.

\end{document}